\documentstyle [12pt,twoside]{article}

\newcommand{\bq}{\begin{equation}}
\newcommand{\ba}{\begin{eqnarray}}
\newcommand{\eq}{\end{equation}}
\newcommand{\ea}{\end{eqnarray}}

\def\bo{{\raise.15ex\hbox{\large$\Box$}}}
\def\bob{{\lower.2ex\hbox{\large$\Box$}}}

\def\TH{{\raise.2ex\hbox{$\displaystyle \bigodot$}\mskip-4.7mu \llap H \;}}
\def\face{{\raise.2ex\hbox{$\displaystyle \bigodot$}\mskip-2.2mu \llap {$\ddot
        \smile$}}}

\def\Hat#1{\rlap{\kern.10em$\widehat{\phantom G}$}#1}
\def\HAt#1{\rlap{\kern.05em$\widehat{\phantom G}$}#1}

\def\cap#1{\rlap{\kern.1em$\widehat{\phantom{G\vrule height.8em}}$}#1{}}
\def\Cap#1{\rlap{\kern.05em$\widehat{\phantom{G\vrule height.8em}}$}#1{}}

\def\leftrightarrowfill{$\mathsurround=0pt \mathord\leftarrow \mkern-6mu
        \cleaders\hbox{$\mkern-2mu \mathord- \mkern-2mu$}\hfill
        \mkern-6mu \mathord\rightarrow$}
\def\overleftrightarrow#1{\vbox{\ialign{##\crcr
        \leftrightarrowfill\crcr\noalign{\kern-1pt\nointerlineskip}
        $\hfil\displaystyle{#1}\hfil$\crcr}}}

\def\frac#1#2{{\textstyle{#1\over\vphantom2\smash{\raise.20ex
        \hbox{$\scriptstyle{#2}$}}}}}

\def\sfrac#1#2{{\vphantom1\smash{\lower.5ex\hbox{\small$#1$}}\over
        \vphantom1\smash{\raise.4ex\hbox{\small$#2$}}}}
\def\bfrac#1#2{{\vphantom1\smash{\lower.5ex\hbox{$#1$}}\over
        \vphantom1\smash{\raise.3ex\hbox{$#2$}}}}
\def\afrac#1#2{{\vphantom1\smash{\lower.5ex\hbox{$#1$}}\over#2}}

\catcode`@=11
\def\underline#1{\relax\ifmmode\@@underline#1\else
        $\@@underline{\hbox{#1}}$\relax\fi}
\catcode`@=12

\def\nis{\nointerlineskip}
\def\Abar{\vbox{\nis\moveright.33em\vbox{
        \hrule width.35em height.04em}\nis\kern.05em\hbox{$A$}}{}}
\def\Dbar{\vbox{\nis\moveright.20em\vbox{
        \hrule width.50em height.04em}\nis\kern.05em\hbox{$D$}}{}}
\def\Gbar{\vbox{\nis\moveright.20em\vbox{
        \hrule width.50em height.04em}\nis\kern.05em\hbox{$G$}}{}}
\def\mbar{\vbox{\nis\moveright.15em\vbox{
        \hrule width.60em height.04em}\nis\kern.05em\hbox{$m$}}{}}
\def\Rbar{\vbox{\nis\moveright.20em\vbox{
        \hrule width.50em height.04em}\nis\kern.05em\hbox{$R$}}{}}
\def\Vbar{\vbox{\nis\moveright.05em\vbox{
        \hrule width.60em height.04em}\nis\kern.05em\hbox{$V$}}{}}
\def\Xbar{\vbox{\nis\moveright.20em\vbox{
        \hrule width.60em height.04em}\nis\kern.05em\hbox{$X$}}{}}
\def\thetabar{\vbox{\nis\moveright.15em\vbox{
        \hrule width.30em height.04em}\nis\kern.05em\hbox{$\theta$}}{}}
\def\Lambdabar{\vbox{\nis\moveright.25em\vbox{
        \hrule width.35em height.04em}\nis\kern.05em\hbox{${\mit\Lambda}$}}{}}
\def\Sigmabar{\vbox{\nis\moveright.25em\vbox{
        \hrule width.50em height.04em}\nis\kern.05em\hbox{${\mit\Sigma}$}}{}}
\def\phibar{\vbox{\nis\moveright.18em\vbox{
        \hrule width.40em height.04em}\nis\kern.05em\hbox{$\phi$}}{}}
\def\chibar{\vbox{\nis\moveright.12em\vbox{
        \hrule width.40em height.04em}\nis\kern.05em\hbox{$\chi$}}{}}
\def\psibar{\vbox{\nis\moveright.23em\vbox{
        \hrule width.40em height.04em}\nis\kern.05em\hbox{$\psi$}}{}}
\def\debar{\vbox{\nis\moveright.18em\vbox{
        \hrule width.35em height.04em}\nis\kern.05em\hbox{$\partial$}}{}}
\def\delbar{\vbox{\nis\moveright.10em\vbox{
        \hrule width.63em height.04em}\nis\kern.05em\hbox{$\nabla$}}{}}

\newskip\humongous \humongous=0pt plus 1000pt minus 1000pt

\newif\ifdtup

\def\begintitle#1#2#3#4
        {\begin{titlepage}
         \centerline{#1 \hfill MdDP-PP-88-#2}
         \begin{center}\vglue .7in
         {\large\bf #3}\\].7in(
         {\bf #4}\\
         {\it Department of Physics and Astronomy}\\
         {\it University of Maryland, College Park, MD 20742}\\].7in(
         {\bf ABSTRACT}\\
         \end{center}
         \begin{quotation}}
\def\endtitle
         {\end{quotation}
          \end{titlepage}
          \newpage}

\begin{document}

\begin{titlepage}

\begin{large}
\par\noindent
an invited plenary talk at: {\it The Seventh Marcel Grossmann Meeting}
\par\noindent
July 1994
\end{large}
\vspace{54pt}
\begin{center}
\begin{Large}

\bf{
CHAOS, REGULARITY, AND NOISE IN
SELF-GRAVITATING SYSTEMS
}

\end{Large}

\vspace{36pt}

\begin{large}
HENRY E. KANDRUP

{\em Department of Astronomy and Department of Physics and}\\
{\em Institute for Fundamental Theory, University of Florida,
Gainesville, FL 32611, USA}
\end{large}

\vspace{54pt}

{\bf Abstract}\\
\end{center}

This paper summarises a number of new, potentially significant, results,
obtained recently by the author and his collaborators, which impact on various
issues related to the gravitational $N$-body problem, both Newtonianly and in
the context of general relativity.

\end{titlepage}

\section{Introduction and motivation}

The overall objective of the research reported herein is the application of
ideas and techniques from modern nonlinear dynamics and nonequilibrium
statistical mechanics to self-gravitating systems, both Newtonianly and in
the context of general relativity, with particular emphasis on the
gravitational $N$-body problem. The basic motivation for this research is a
desire to identify some of the physical processes which can play a role in
determining the structure and evolution of self-gravitating systems. The
results described here will, for specificity, typically be formulated in the
language of galactic dynamics. However, it should be evident that they
also have potential implications in other settings as well, including, e.g.,
statistical quantum field theory in the early Universe.

It should be stressed that the principal focus here is {\it not} on the
detailed modeling of any specific class of astronomical object where, in
particular, other nongravitational effects, such as dissipative hydrodynamics,
can be important. However, the results reported here should find relatively
direct applications to the study of systems like elliptical and lenticular
galaxies, which are known to be gas poor, albeit not as gas poor as they were
ten years ago.

In approaching the study of self-gravitating systems, there are several
different approaches which one might adopt. At the most fundamental level, one
can attack the full $N$-body problem, either by performing and analysing
numerical simulations or by proving (hopefully useful) mathematical theorems
which provide insights into the qualitative character of the evolution. In
either case, the focus here is {\it not} on solar system type problems,
involving a relatively small number of masses, one or two of which are much
larger than the others. Rather, the principal focus is on collections of large
numbers $N$ of objects, comparable in mass, in particular the problem of the
$N \to\infty$ limit.

Conventional wisdom says that, in this large $N$ limit, such a collection of
comparable masses can be described by a collisionless Boltzmann equation, i.e.,
what the mathematician would term the Vlasov-Poisson or Vlasov-Einstein system.
Such a description involves the assumption that, Newtonianly, particles follow
trajectories in the self-consistent gravitational potential associated with the
average mass distribution. Relativistically, one supposes that the particles
follow geodesics in a spacetime, the form of which is determined by the stress
energy tensor associated with the average mass distribution. In this general
context, two things need to be done, namely (1) to determine precisely the
conditions under which such a mean field description is justified and then (2)
to understand the qualitative and quantitative implications of this
description.

In this connection, it is important to stress that, although the Vlasov-Poisson
system was first formulated by Jeans in the context of galactic
dynamics\cite{J}
more than twenty years before it was formulated in plasma physics,\cite{Vl}
the gravitational equation is substantially less well motivated than
is the plasma analogue. In particular, the absence of shielding prevents a
systematic derivation except for the special case of cosmological systems,
which are nearly homogeneous and of finite age.\cite{BKS}
The problem becomes especially acute for the case of
relativistic systems, the point being that the derivation of the corresponding
Vlasov-Maxwell system relies heavily on the linearity of Maxwell's
equations,\cite{Kl} whereas Einstein's equation is nonlinear.

A yet simpler tact involves the consideration of orbits in a fixed potential.
The idea here is to specify one's favourite potential, chosen to represent
the bulk potential of some physical object, to study in detail the form of
orbits in this potential, and only at the end of the day to incorporate the
fact that the potential must be determined self-consistently by the mass
distribution associated with the orbits themselves.

If one chooses to focus simply on an average potential, in the context of a
Vlasov description or otherwise, there remains the important question of
determining precisely, both qualitatively and quantitatively, what sorts of
effects have been ignored. In other words, there remains the problem of
{\it structural stability.} Only to the extent that these additional effects
have been appropriately identified and understood can one say that one has
a satisfactory understanding of so-called ``collisionless stellar dynamics.''

The aim of this talk is to illustrate each of the preceding aspects by
explaining several new results that have been derived during the past three
or four years. Section $2$ focuses on the full Newtonian $N$-body
problem. Section $3$ then turns to the collisionless Boltzmann equation.
Section $4$ addresses several issues related to the problem of orbits in a
fixed potential, and Section $5$ concludes with a discussion of some aspects
of the problem of structural stability.

\section{The gravitational $N$-body problem}

Detailed numerical simulations over the past thirty years have established
that, given generic initial conditions corresponding to a bound configuration,
a self-gravita-ting system of point masses will typically exhibit a rapid
approach, on a characteristic crossing time $t_{cr}$, towards a statistical
quasi-equilibrium where, in a time-averaged sense, the system only shows
subsequent systematic variability on substantially longer timescales. Moreover,
there has evolved a substantial and detailed conventional wisdom which serves,
at least roughly, to determine what kinds of initial data give rise to what
kinds of final quasi-equilibria. However, despite these impressive successes,
one does not completely understand this process. Indeed, at a truly fundamental
level there is no real understanding of {\it why this happens.}

To obtain some insights into this basic question, it is natural to identify
various microscopic and mesoscopic phenomena which could perhaps conspire to
produce the qualitative macroscopic evolution that is observed in numerical
experiments. The aim here is to identify two such phenomena.
\par\noindent
{\it 1. Viewed microscopically in the many particle phase space, Newtonian
$N$-body simulations exhibit an exponentially sensitive dependence on the
specific choice of initial conditions}. Specify unperturbed initial data,
$\{{\bf r}^{u}_{A}(0)$ ${\bf p}^{u}_{A}(0)\}$, $(A=1,...,N)$, and perform a
simulation. Then specify perturbed initial data,
$\{{\bf r}^{p}_{A}(0)$ ${\bf p}^{p}_{A}(0)\}$, and repeat. What one discovers
thereby is that, generically, quantities like the total $N$-particle
configuration space perturbation
$$\sum_{A=1}^{N} |{\delta}{\bf r}_{A}(t)|^{2}{\;}{\equiv}{\;}
\sum_{A=1}^{N} |{\bf r}^{p}_{A}(t)-{\bf r}^{u}_{A}(t)|^{2} \eqno(1) $$
will typically grow exponentially in time $t$ on a relatively short time
scale, even though the unperturbed and perturbed evolution are essentially
identical at the macroscopic level.

This fact was first observed by S. Ulam in the early 1960's, and the classic
paper on the subject is by Miller.\cite{Mi} However, only in the last several
years, with the advent of improved computers, has this instability been
studied systematically in complete and gory
detail.\cite{KS91,KS92,KSW,GHH,KMS94}

The net result of these investigations is that this instability is an
exceedingly robust phenomenon, which proceeds generically on a characteristic
crossing time $t_{cr}$, independent of many/most details. In particular, the
timescale associated with this instability is independent of the detailed
choice of initial data and the detailed choice of the initial perturbations,
as well as the specific diagnostics used to quantify the growth of the
perturbation -- e.g., configuration or momentum space perturbation, the total
$N$-particle perturbation or the perturbation of ``typical'' or
``representative'' particles. The timescale is also insensitive to a possible
distribution of masses, provided that everything is not dominated by a few
particularly massive particles.

More significantly, the simulations also suggest strongly that the rate is
insensitive to the total particle number $N$, provided at least that
$N{\;}{\gg}{\;}2$. Thus, e.g., for $200{\;}{\le}{\;}N{\;}{\le}{\;}4000$
one observes no appreciable changes if everything is scaled in terms of an
$N$-dependent characteristic time $t_{cr}$. In particular, there is no sense in
which the instability appears to ``turn off'' in the limit of large $N$.
Indeed, Goodman {\it et al}\cite{GHH} have predicted that the instability
should accelerate for large $N$, with the characteristic timescale $t_{*}$
scaling as $t_{*}{\;}{\sim}{\;}t_{cr}/{\rm ln}{\,}N$. Interestingly, $t_{cr}$
is the same timescale on which generic initial data evolve towards a
macroscopic quasi-equilibrium.
\par\noindent
{\it 2. Gravitational $N$-body simulations evidence a considerable ``memory.''}
Viewed microscopically or mesoscopically, the quasi-equilibrium does {\it not}
correspond to a particularly well-shuffled state. Many aspects of the initial
data for individual particles are forgotten, but other aspects are in fact
remembered. The strongest correlation between initial and final conditions,
from which all others may perhaps derive, is between the initial and final
values of the binding energy. Both for {\it isolated systems} approaching a
quasi-equilibrium and for {\it collisions} between pairs of objects, e.g.,
spherical polytropes and other axisymmetric or triaxial near-equilibria, there
is a strong correlation between the initial and final values of the binding
energy.

Specifically, one observes that particles initially with high binding
energy tend to end up with high binding energy, low with low, and intermediate
with intermediate, even for an evolution that involves rapid, violent changes
in the bulk potential, so that there is no obvious sense in which the binding
energy should behave as an adiabatic invariant.\cite{vA,QZ,FME,KMS93}
This phenomenon can be quantified at a coarse-grained mesoscopic level,
through various binnings of the orbital data.\cite{FME,KMS93} Alternatively,
it can be quantified at the completely microscopic level through the
computation of a rank correlation between the initial and final
{\it orderings} of particles in terms of their binding energies.\cite{KMS93}
Such a computation shows that there exist strong,
albeit not complete, correlations between initial and final conditions.
The absence of a complete correlation is at least partly ``collisional'' in
origin, but appears to persist even in the limit of large $N$, where,
according to conventional wisdom, the system should be essentially
collisionless in character.

To summarise: In the gravitational $N$-body problem one is confronted with a
system that exhibits a rapid
macroscopic evolution towards a statistical quasi-equilibrium. Viewed
microscopically, this evolution is characterised by an exponentially sensitive
dependence on the specific choice of initial conditions. However, despite this
sensitive dependence, the quasi-equilibrium does not correspond, either
microscopically or mesoscopically, to a particularly well-shuffled state.

\section{ The collisionless Boltzmann equation}

The principal message of this section is that {\it the collisionless Boltzmann
equation}, i.e., a mean field Vlasov description, {\it is a constrained
Hamiltonian system}. This fact was first established by Morrison\cite{Mo} for
the electrostatic Vlasov-Poisson system, and is equally true for the
analogous gravitational Vlasov-Poisson system. In this setting, the
fundamental dynamical variable is the one-particle distribution function,
$f({\bf x},{\bf p},m)$, evaluated at a fixed time $t$, which is interpreted
as a number density of particles of mass $m$ at
the spatial point ${\bf x}$ with conjugate momentum ${\bf p}$. The
gravitational potential ${\Phi}(t)$ is viewed as a functional of $f(t)$,
constructed in terms of an appropriate Green function. The phase space
is then the infinite-dimensional space of distribution functions. The
Hamiltonian character of
the evolution is manifest through the identification of a Hamiltonian function
$H[f]$ and a cosymplectic structure $\{\;.\;,\;.\;\}$, in terms of which the
Vlasov equation takes the form ${\partial}f/{\partial}t=\{f,H\}$.

A generalisation to the spherically symmetric Vlasov-Einstein system is
completely straightforward.\cite{KMo} The crucial point
is that, for spherical systems, the gravitational degrees of freedom are
not triggered: given appropriate boundary conditions,
one can view the spacetime metric $g_{ab}$ at any given time $t$ as a
functional of the distribution function $f$ at that same $t$.

The full Vlasov-Einstein system is substantially more complicated, since, in
the general case, one must incorporate the field degrees of freedom. However,
the analysis still turns out to be straightforward, at least in
principle.\cite{KO94}  The basic formulation is analogous
to the Hamiltonian formulation of the Vlasov-Maxwell system,\cite{MW,KO93b}
the nonlinearity of the Einstein equation not playing a significant role.

Working in the context of the {\it ADM}
formulation of general relativity, there are now three different dynamical
variables, each defined on $t={\rm const}$ hypersurfaces, namely (1) the
distribution function, $f({\bf x},{\bf p},m)$, (2) the spatial three-metric,
$h_{ab}({\bf x})$, and (3) the conjugate field momentum, ${\Pi}^{ab}({\bf x})$.
The natural phase space is the infinite-dimensional space coordinatised by
these three variables.

In this case the cosymplectic structure given as the sum of two pieces, namely
(1) the functional Poisson bracket of vacuum gravity and (2) the matter
bracket appropriate for the spherically symmetric Vlasov-Einstein system
(which coincides also with the bracket for the Vlasov-Poisson system).
Explicitly, for two functions $F[f,h_{ab},{\Pi}^{ab}]$ and
$G[f,h_{ab},{\Pi}^{ab}]$,
$${\langle}F,G{\rangle}=16{\pi}{\int}{\,}d^{3}x{\,}
{\Biggl(}{{\delta}F\over {\delta}h_{ab}}{{\delta}G\over {\delta}{\Pi}^{ab}}-
{{\delta}G\over {\delta}h_{ab}}{{\delta}F\over {\delta}{\Pi}^{ab}}
{\Biggr)}
+{\int}{\,}d^{3}xd^{3}pdm{\,}f{\,}
{\Biggl[}{{\delta}F\over {\delta}f},{{\delta}G\over {\delta}f}{\Biggr]},
\eqno(2) $$
where ${\delta}/{\delta}X$ denotes a functional derivative with respect to
the variable $X$ and
$$[f,g]={{\partial}f\over {\partial}x^{i}}{{\partial}g\over {\partial}p_{i}}-
{{\partial}g\over {\partial}x^{i}}{{\partial}f\over {\partial}p_{i}}
\eqno(3)$$
denotes an ordinary three-dimensional Poisson bracket.

To give meaning to variations ${\delta}X$, one requires a rule identifying
particle coordinates $\{x^{a},p_{a}\}$  and $\{x'^{a},p'_{a}\}$ in two
nearby cotangent bundles. In the context of this $3+1$ formulation, it is
natural to identify spatial coordinates and conjugate momenta, as well as time
$t$ and mass $m$, i.e.,
$${\bf x}'={\bf x} \qquad {\bf p}'={\bf p} \qquad t'=t \qquad
m'=m. \eqno(4)$$
However, other choices are also possible.\cite{IT,IK}

The Hamiltonian $H=H_{G}+H_{M}={\int}d^{3}x{\cal H}_{G}+
{\int}d^{3}x{\cal H}_{M}$ is also given as the sum
of two pieces, namely (1) the {\it ADM} Hamiltonian $H_{G}$ of vacuum
gravity, i.e.,
$${\cal H}_{G}={1\over 16{\pi}}
h^{1/2}{\Biggl\{}N{\Bigl[}-^{(3)}R+h^{-1}{\bigl(}{\Pi}^{ab}{\Pi}_{ab}
-{1\over 2}{\Pi}^{2}{\bigr)}{\Bigr]}$$
$$- 2N_{b}{\bigl[}D_{a}(h^{-1/2}{\Pi}^{ab}){\bigr]}
+2D_{a}(h^{-1/2}N_{b}{\Pi}^{ab}){\Biggr\}} \eqno(5)$$
and (2) a matter Hamiltonian $H_{M}$ constructed as the integral of the local
energy density, i.e.,
$$H_{M}={\int}d{\Gamma}{\,}f{\cal E}={\int}{\;}Nh^{1/2}d^3x{\,}T^{t}_{\;t}.
\eqno(6)$$
Here $d{\Gamma}=d^{3}xd^{3}p{\,}dm$ denotes the covariant seven-dimensional
volume element on a $t={\rm const}$ hypersurface, $D_{a}$ a covariant
derivative on the hypersurface, and ${\cal E}({\bf x},{\bf p},m)$ $=|p_{t}|$
the particle energy. $N$ and $N_{a}$ correspond respectively to the lapse
function and shift vector.

This formulation reproduces the Vlasov-Einstein system in the sense that the
equation  ${\partial}F/{\partial}t={\langle}F,{\cal H}{\rangle}$
for arbitrary functions $F[f,h_{ab},{\Pi}^{ab}]$ is equivalent to the
Vlasov-Einstein system in its usual $3+1$ form: The distribution function
$f$ satisfies
${{\partial}f/{\partial}t}=[{\cal E},f], $
and ${\partial}h_{ab}/{\partial}t$ and ${\partial}{\Pi}^{ab}/{\partial}t$
both satisfy the appropriate $3+1$ Einstein equations sourced by
$T^{a}_{\;\;b}[f]$. For the spherically symmetric case, with the metric viewed
as a functional of $f$, the first term in the bracket ${\langle}F,G{\rangle}$
disappears and the Hamiltonian $H$ reduces to the {\it ADM} mass, $H_{ADM}$,
realised as a volume integral in the natural fashion facilitated by
Schwarzschild coordinates.

Such a Hamiltonian formulation of the collisionless Boltzmann equation is in
fact an example of a much more general result. Specifically, consider {\it any}
Hamiltonian theory for a system with multiple degrees of freedom, and
then construct the associated mean field description appropriate in the limit
that correlations amongst the degrees of freedom are negligible (i.e., a
statistical description in which the full $N$-``particle'' distribution
function is approximated as factorising into a product of reduced
one-``particle'' distribution functions). There is then a precise mathematical
sense in which {\it the mean field description of this Hamiltonian system
is itself Hamiltonian}.\cite{K94}

The fact that the collisionless Boltzmann equation, or any other mean field
theory, is Hamiltonian is significant in that a Hamiltonian evolution is much
more restricted than a non-Hamiltonian evolution. However, of more pragmatic
importance perhaps is the fact that this Hamiltonian formulation permits, for
the first time, a clear geometric approach to the problem of stability for
general equilibrium solutions to the Vlasov-Einstein system. Here an
``equilibrium solution'' $\{f_{0},h_{ab}^{0},{\Pi}^{ab}_{0}\}$ entails a
stationary matter distribution, corresponding to a spacetime that admits a
timelike Killing field.

The crucial fact facilitating this geometric approach is that every
such equilibrium is an energy extremal, so that the first variation
${\delta}^{(1)}H$ vanishes identically for perturbations
$\{{\delta}f,{\delta}h_{ab},{\delta}{\Pi}^{ab}\}$ which satisfy the
constraints. The field constraints are enforced by restricting attention to
perturbed initial data for which ${\delta}H/{\delta}N=
{\delta}H/{\delta}N^{a}=0$. The matter constraints, a reflection of Liouville's
Theorem, imply that the perturbed distribution function must be generated
from the unperturbed $f_{0}$ via a canonical transformation in terms of some
generating function $h$, i.e., $f_{0}+{\delta}f={\rm exp}([h,\,.\,])f_{0}=
f_{0}+[h,f_{0}]+ {1\over 2}{\Bigl[}h,[h,f_{0}]{\Bigr]}+ ....$.

The fact that the equilibrium is an energy extremal implies that stability
hinges on the sign of the second variation ${\delta}^{(2)}H$. Indeed,
modulo infinite-dimensional technicalities the situation is analogous to the
problem of stability for mechanical Hamiltonian systems. If
${\delta}^{(2)}H>0$ for all infinitesimal perturbations, linear
stability is guaranteed. Alternatively, if ${\delta}^{(2)}H$ is
indeterminate, one cannot necessarily infer a linear instability, but one
{\it does} expect at least nonlinear instability and/or instability in the
presence of dissipation.\cite{BKMR}

Indeed, one can actually prove that {\it generic rotating axisymmetric
equilibria are always unstable towards dissipation}, as provided, e.g., by the
emission of gravitational radiation.\cite{K91} This is the
collisionless analogue of the theorem\cite{FS} that all
rotating perfect fluid stars are unstable. Neither for collisionless nor
collisional systems is there any guarantee that the timescale associated with
this instability is sufficiently short to be of interest astronomically.
However, it {\it is} significant that general relativity triggers a generic
instability which, apparently, is insensitive to the form of the
self-gravitating matter. The astronomical implications of this instability
are currently under investigation.

This general approach to stability can also be adapted to the consideration
of steady-state equilibria, such as an homogeneous and isotropic Friedman
cosmology, where it provides an interesting derivation of the Jeans
instability.\cite{KO93a} Viewed Newtonianly, such an expanding Universe
corresponds in the ``inertial'' frame to a system characterised by a
time-independent Hamiltonian which finds itself in a time-dependent steady
state. This explicit time-dependence can be removed by effecting a
time-dependent canonical transformation into the average ``comoving'' frame.
This transformation leads to a new time-dependent Hamiltonian $H(t)$. However,
the first variation ${\delta}^{(1)}H$ vanishes identically, and the second
variation ${\delta}^{(2)}H$ can be shown to satisfy
$d{\delta}^{(2)}H/dt{\;}{\le}{\;}0$. A simple Liapounov argument therefore
guarantees that the existence of negative energy perturbations
${\delta}^{(2)}H<0$ for
sufficiently long wavelengths must imply an instability: If the system be
perturbed in such a fashion that ${\delta}^{(2)}H<0$, the energy can only
become more negative, so that the ``distance'' from equilibrium, as probed by
the magnitude of ${\delta}^{(2)}H$, can only increase.

\section{ Transient Ensemble Dynamics}

In studying the properties of orbits in a fixed potential, it is natural
to apply the technology of nonlinear dynamics, as has been developed over
the past several decades. However, in many settings involving gravitational
systems, the utilisation of this technology may require a new twist. The
standard formulation of nonlinear dynamics typically involves a theory
of {\it asymptotic orbital dynamics}, which focuses primarily, if not
exclusively, on the long time behaviour of individual orbits. However, for
many astronomical systems this may not be appropriate.

For example, in terms of their natural
timescale, galaxies are relatively young objects, only ${\sim}{\;}100-200$
crossing times $t_{cr}$ in age, so that it is not at all obvious that an
asymptotic $t\to\infty$ limit is well motivated physically. Thus, e.g.,
standard estimates of Liapounov exponents typically require
integrations for times $t{\;}{\ge}{\;}10^{4}t_{cr}$, a period that is
orders of magnitude longer than the age of the Universe, $t_{H}$.
Moreover, it is arguably true that, in many astronomical systems, individual
orbits are not the fundamental objects of interest. It is, for example, obvious
that one cannot track individual orbits of stars within a galaxy. All that one
can detect are properties like the overall brightness distribution which
reflect the contributions of many stars. Similarly, it is evident that one must
focus on collections of orbits if he or she wishes ultimately to address
the problem of self-consistency.

For these reasons, it would seem more natural to consider instead a theory of
{\it transient ensemble dynamics}, which focuses on the statistical
properties of ensembles of orbits, restricting attention exclusively to short
timescales, $t<t_{H}$, and recognising that much of the observed behaviour
may be intrinsically transient in character.

This distinction may  not be all that important for integrable, or
near-integrable potentials, which contain only regular orbits. However, there
is no reason to assume that the bulk potential associated with a
self-gravitating system is integrable, or even near-integrable, and there
{\it are} indications from numerical experiments that objects like rotating
elliptical galaxies and barred spiral galaxies may admit large numbers of
stochastic, or chaotic, orbits.\cite{SS,PF}

Roughly, {\it regular orbits} correspond to orbits that have regular shapes
and are characterised by simple topologies, e.g., box orbits in two dimensions
with trajectories that resemble a Lissajous figure or tube orbits in three
dimensions that are restricted to a region with the topology of a torus. By
contrast, {\it stochastic orbits} are manifestly irregular in shape and
appear to ``run all over'' phase space. Unlike regular orbits, stochastic
orbits exhibit an exponentially sensitive dependence on initial conditions, as
manifest by the fact that such orbits are characterised by a positive Liapounov
exponent.\cite{Ch79}

The crucial point to be illustrated below is that, at least when considering
stochastic orbits, the transient ensemble perspective can prove extremely
important.

Consider, for example, the scattering of photons incident on a multi-black
hole system; or similarly, consider a star moving in a nonspherical potential,
supposing that the star is unbound but that, owing to the shape of the
potential, it can only escape in certain directions. For each of these
problems, one discovers that, in certain phase space regions, the direction
and time of escape to infinity exhibits a very sensitive dependence on initial
conditions, in fact a fractal dependence, this being an example of what the
nonlinear dynamicist would call {\it chaotic scattering}.\cite{Sm}

Naively, one might anticipate that this complex microscopic behaviour would
lead to an equally complex description when considering the evolution of
ensembles of orbits. However, this is {\it not} always the case. At least in
certain cases, one observes instead striking regularities, which lead to a
simple scaling behaviour,\cite{CKK} that may actually be universal.\cite{SCK}
{\it The very fact that the microscopic evolution is complex appears to be
responsible for the fact that the macroscopic evolution is very simple.}

As a concrete example, consider orbits in the two-dimensional potential
$V(x,y)={1\over 2}(x^{2}+y^{2})-{\epsilon}x^{2}y^{2},$
holding fixed the value of the energy $E$ and studying as a function of
${\epsilon}$ the evolution of orbits initially localised in a small phase space
region. For ${\epsilon}$ below a critical value ${\epsilon}_{1}=1/(4E)$,
escape is impossible energetically. For values of ${\epsilon}$ slightly above
${\epsilon}_{1}$, escape {\it is} possible energetically, but only very few
particles escape on short timescales and the time of escape exhibits no
striking regularities, except that the escape probability eventually appears
to decay towards zero. However, for ${\epsilon}$ above another critical
value ${\epsilon}_{2}$ (only determined numerically), one sees the onset
of striking scaling behaviour:
\par\noindent
1) After the decay of initial transients, the escape probability per unit time
approaches a constant value $P_{\infty}({\epsilon})$, which is independent of
initial conditions, i.e., the location of the phase space region from which
the initial ensemble was chosen. Moreover, this rate scales as
$P_{\infty}({\epsilon}){\;}{\sim}{\;}({\epsilon}-{\epsilon}_{2})^{\alpha}$
for a critical exponent ${\alpha}$.
\par\noindent
2) For fixed size of the initial phase space region probed by the ensemble,
the time $T_{\infty}$ required to converge to $P_{\infty}$ also depends on
${\epsilon}$ and satisfies
$T_{\infty}({\epsilon}){\;}{\sim}{\;}({\epsilon}-{\epsilon}_{2})^{-\beta}$.
\par\noindent
3) For fixed ${\epsilon}$, the convergence time $T_{\infty}$ depends on the
linear size $R$ of the phase space region that was probed initially, satisfying
$T_{\infty}(R){\;}{\sim}{\;}R^{-\delta}$
\par\noindent
Moreover, to within statistical errors
${\alpha}-{\beta}-{\delta}{\;}{\approx}{\;}0$. In a certain sense, the
qualitative change in behaviour at ${\epsilon}={\epsilon}_{2}$ is like a
phase transition, complete with a critical ``slowing down'' as the ``order
parameter'' ${\epsilon}-{\epsilon}_{2}\to 0$.

As another example in which the transient ensemble perspective is important,
consider the behaviour of ensembles of orbits of fixed energy $E$, evolving in
a two-dimensional time-independent potential $V(x,y)$, which admits both
regular and sto-chastic orbits. If $V(x,y)$ is bounded from below and diverges
at infinity, the constant energy hypersurfaces will be compact, so that the
notion of ``equilibrium'' is well defined. One might therefore anticipate that
generic ensembles of initial conditions will evolve towards an {\it invariant
distribution} corresponding to a statistical equilibrium.

To test this hypothesis, one can select localised ensembles of initial
conditions of fixed $E$, corresponding to stochastic orbits initially located
far from any regular phase space regions, and then evolve these initial data
into the future. At least for certain potentials,\cite{KM94a,MABK} one then
observes
that the orbits do indeed disperse in such a fashion as to exhibit a
coarse-grained evolution towards a quasi-equilibrium, which is at least
approximately
time-independent. This approach is, moreover, exponential in time and
characterised by a rate ${\Lambda}(E)$ which is independent of the specific
choice of initial data. The characteristic timescale $t_{*}={\Lambda}^{-1}$
is typically ${\ll}{\;}100t_{cr}$, so that, in ``physical units'' for a galaxy,
$t_{*}{\;}{\ll}{\;}t_{H}$.

One also observes that the rate ${\Lambda}(E)$ is comparable in magnitude to
the value of the Liapounov exponent ${\chi}(E)$, which\cite{BGS}
probes the average rate of instability exhibited by orbits
of energy $E$. There is, moreover, a direct correlation between ${\Lambda}$
and ${\chi}$ in the sense that, e.g., both have similar curvatures when viewed
as functions of $E$. This is particularly tantalising in view of
the fact that, for the $N$-body problem, the timescale associated with the
approach towards a statistical quasi-equilibrium is comparable in magnitude to
the timescale associated with the instability towards small changes in initial
conditions.

Despite these regularities, there is no guarantee that this apparent
equilibrium coincides with the true invariant distribution, and, in general,
it will {\it not}! Astronomers are well acquainted with the fact that
collisionless equilibria do not constitute true $N$-body equilibria, which
must incorporate discreteness effects that become important on sufficiently
long timescales. However, the key point here is different, and more
fundamental: {\it Even motion in a smooth two-dimensional potential may
not yield a uniform approach towards a true statistical
equilibrium}.\cite{MABK,KM94b}

The explanation of the discrepancy between the true and approximate equilibria
is simple. Viewed over sufficiently long timescales, there are only two
different classes of orbits, namely regular orbits, with vanishing Liapounov
exponent ${\chi}$, and stochastic orbits, for which ${\chi}>0$. The distinction
between these two classes is, moreover, absolute, since members of the two
different classes are separated by invariant {\it KAM} tori. If, e.g., one
were to compute a surface of section, plotting coordinates $x$ and $p_{x}$
for successive intersections of the $y=0$ hyperplane, he or she would
generically find islands of regularity embedded in a surrounding stochastic
sea.

However, this is not the whole story. Lurking in the shallows of the stochastic
sea, slightly away from the shore, are cantori,\cite{Ma} these
corresponding to fractured {\it KAM} tori, associated with the breakdown of
integrability, which contain a cantor set of holes. The point then is that
these cantori serve as partial barriers that divide the stochastic orbits into
two subclasses, namely confined, or {\it sticky}, stochastic orbits which are
trapped near the regular islands, and unconfined, or {\it filling}, stochastic
orbits which travel unimpeded throughout the rest of the stochastic sea.

Because of the holes in the cantori, these barriers are not absolute, so that
orbits can in fact change from one class to another via so-called {\it
intrinsic diffusion}. However, this process is a slow one, requiring orbits
to wend their way through a maze (cf. the ``turnstile model'' of MacKay,
Meiss, and Percival,\cite{MMP} so that the characteristic timescale is
typically ${\gg}{\;}100t_{cr}$, i.e., much longer than the age of the Universe.

What this implies is that, on short times, ensembles of orbits initially
outside the cantori will evolve towards a {\it near-invariant distribution}
which uniformly populates the filling regions, but avoids the confined regions.
The situation is analogous to the classical effusion problem. Consider two
evacuated cavities connected one with another by an extremely narrow conduit,
and suppose that gas is inserted into one of the cavities. If the conduit be
sufficiently narrow, the timescale on which gas effuses from one cavity to
the other will be much longer than the timescale on which the gas spreads to
fill the original cavity. This implies, however, that, even though the true
equilibrium corresponds to a uniform density concentration throughout both
cavities, one can speak meaningful of a shorter time quasi-equilibrium, in
which the original cavity is populated uniformly and the other is essentially
empty.

Significantly, these two different populations of stochastic orbits are
fundamentally dissimilar in terms of their stability properties as well as
where they are located in phase space. Although both sticky and filling
stochastic orbits are exponentially unstable, there is a precise sense in
which the sticky orbits are less unstable overall than are the filling orbits.
Specifically, if one computes {\it local Liapounov exponents},\cite{GBP}
${\chi}({\Delta}t)$, for different
ensembles of stochastic orbits, integrating for some relatively short interval
${\Delta}t$, he or she will find\cite{MABK,KM94b} that
the typical ${\chi}({\Delta}t)$ for a sticky orbit is substantially
smaller than the typical ${\chi}({\Delta}t)$ for a filling orbit. Indeed,
the composite {\it distribution of local Liapounov exponents} (i.e.,
distribution of instability timescales) generated from a sampling of the true
invariant measure appears to be given, at least approximately, as a sum of two
different near-Gaussian distributions with unequal means.

It should perhaps be noted explicitly that the general conclusions recounted
in this Section have been observed for several different potentials, with
rather different symmetries, including (1) the dihedral $D4$ potential of
Armbruster, Guckenheimer, and Kim,\cite{AGK} (2) the sixth order truncation of
the three-particle Toda\cite{To} lattice potential, and (3) a generalised
anisotropic Kepler potential of the form
$$V(x,y)=-{1\over {\Bigl(}1+x^{2}+y^{2}{\Bigr)}^{1/2}}
-{m\over {\Bigl(}1+x^{2}+ay^{2}{\Bigr)}^{1/2}}, \eqno(7)$$
with constant $m$ and $a$, for $E<0$. The fact that these diverse potentials,
which are fundamentally different in appearance, yield similar conclusions,
both qualitatively and semi-quantitatively, would suggest strongly that these
conclusions are robust, depending only on such topological features as the
existence of {\it KAM} tori and cantori.

The existence of confined stochastic orbits is of potential importance
astronomically because such orbits can help (the theorist) support various
sorts of structures, e.g., bars in a spiral galaxy. It is natural to assume
that, in systems like galaxies, regular orbits serve to provide the
skeleton to support various structures. However, because of resonance overlap
one may find that, near corotation and other resonances, the desired regular
orbits do not exist, even though sticky stochastic orbits are present.

Finally, it should be stressed that one can observe similar short time ``zones
of avoidance'' in higher dimensional systems as well. The key point physically
is that {\it just because a region of phase space is connected, so that orbits
can pass throughout the entire region, does not mean that all of the region
will be accessed on comparable timescales.}

\section{ Structural stability of the smooth potential approximation}

The collisionless Boltzmann equation is a Hamiltonian system which neglects
various realistic non-Hamiltonian irregularities that must be present in any
self-gravitating system. One obvious point is that such a Vlasov description
neglects entirely all discreteness effects, i.e., ``collisions,'' by
idealising the system as a continuum, rather than a collection of nearly point
mass objects. Viewed in the $N$-particle phase space, the statistical
description of an isolated $N$-body evolution is of course Hamiltonian.
However, when projected into the reduced one-particle phase space, any
allowance for particle-particle correlations that transcend a mean field
description necessarily breaks the Hamiltonian constraints.\cite{K94}
Another point, perhaps less obvious but equally important, is that a Vlasov
description also neglects any couplings to an external environment. In the
past, astronomers have been wont oftentimes to pretend that galaxies exist in
splendid isolation but, over the past several decades, it has become
increasing evident that such an approximation may not be justified.\cite{ZW}

Detailed modeling of these sorts of perturbing influences may prove extremely
complex. In particular, an external environment can give rise to a variety of
different effects characterised by a broad range of timescales. Those
influences proceeding on timescales ${\sim}{\;}t_{cr}$ will be particularly
complicated, in that the details of their effects may depend very sensitively
on the details of the environment. However, there is a well-established
paradigm in statistical physics,\cite{Ch43,KTH}
dating back to the beginning of the century,\cite{E}
which would suggest that irregularities proceeding on shorter timescales,
${\ll}{\;}t_{cr}$, can oftentimes be modelled as friction and noise, related
via a fluctuation-dissipation theorem. This idea underlies, for example,
Chandrasekhar's original formulation\cite{Ch42} of so-called ``collisional
stellar dynamics.''

It is therefore natural to investigate the {\it structural stability}
of Hamiltonian trajectories towards the effects of friction and noise.
This was done\cite{KM94c,HKM94a} by effecting large numbers of {\it Langevin
simulations},
in which the deterministic equations of motion were perturbed by allowing
for (1) a dynamical friction $-{\eta}{\bf p}$, which serves
systematically to remove energy from the orbits and (2) random kicks, modeled
as white noise with temperature, or mean squared velocity,
${\Theta}$, which serve systematically to pump energy back into the orbits.
As a first simple test, ${\eta}$ was assumed to be constant, in which case
the fluctuation-dissipation theorem implies that the noise must be
additive, rather than multiplicative.\cite{LS,HK}

Thus, in units with particle mass $m=1$, one is led explicitly to equations of
motion of the form
$${d{\bf r}\over dt}={\bf p} \qquad {\rm and} \qquad
{d{\bf p}\over dt}=-{\nabla}V({\bf r})-{\eta}{\bf p}+{\bf F},  \eqno(8) $$
where ${\bf F}$ is characterised completely by its statistical properties.
Here, e.g., component by component,
$${\langle}F_{i}(t){\rangle}=0 \hskip .2in {\rm and} \hskip .2in
{\langle}F_i(t_1)F_j(t_2){\rangle}=
2{\Theta}{\eta}{\delta}_{ij}{\delta}_{D}(t_1-t_2), \eqno(9)$$
where the angular brackets denote a statistical average. The idea is to effect
large numbers of different realisations of the same initial conditions, and to
analyse these realisations to extract statistical properties.

It is well known that even very weak friction and noise will eventually become
important on sufficiently long timescales. In particular, one knows that, on
the natural timescale $t_{R}{\;}{\sim}{\;}{\eta}^{-1}$,
these effects will try to force the system to evolve towards a thermal state.
The question of relevance here is quite different: {\it Can the friction and
noise have substantial effects already on much shorter timescales}
${\ll}{\;}t_{R}?$ The conventional wisdom in astronomy is that the answer to
this is: {\it no}! For example, the standard assumption that ``collisionality''
is irrelevant in a galaxy relies completely on the observation that the
natural timescale associated with binary encounters is much longer than the
age of the Universe.\cite{Ch42}

The Langevin simulations were effected for total times
$t{\;}{\le}{\;}200t_{cr}$, and involved friction and noise corresponding to
a broad range of characteristic timescales,
$10^{3}{\;}{\le}$ $t_{R}/t_{cr}{\;}{\le}{\;}10^{12}$. The most significant
conclusions derived from these simulations are the following:

When viewed in terms of the {\it collisionless invariants}, i.e., the
quantities that are conserved in the absence of the friction and noise,
these perturbing influences only serve to induce a classical {\it diffusion
process}, with the unperturbed and perturbed orbits diverging significantly
only on a timescale $t_{R}{\;}{\sim}{\;}{\eta}^{-1}$. Thus, in particular,
$${\delta}E_{rms}^{2}{\;}{\equiv}{\;}{\langle}|E_{unp}-E_{per}|^{2}{\rangle}=
A(E)E{\Theta}{\eta}t, \eqno(10) $$
where $A(E)$ is a slowly varying function of $E$ with magnitude of order
unity. In this sense, the conventional wisdom is confirmed.

However, when viewed in {\it configuration space} or {\it momentum space},
the effects are more complicated, and actually depend on orbit class.
Unperturbed and perturbed {\it regular} orbits only diverge as
a power law in time, i.e.,
${\delta}r_{rms},{\delta}p_{rms}{\;}{\sim}{\;}t^{p}$, so
that, once again, one only gets macroscopic deviations after a time
$t_{R}{\;}{\sim}{\;}{\eta}^{-1}$. However, unperturbed and perturbed
{\it stochastic} orbits diverge exponentially at a rate set by the Liapounov
exponent ${\chi}$, so that, even for very weak friction and noise, one gets
macroscopic deviations within a few crossing times. In particular, when
considering ensembles of stochastic initial conditions, one
observes a simple scaling
$${\delta}r_{rms},{\delta}p_{rms}{\;}{\sim}{\;}({\Theta}{\eta})^{1/2}
{\rm exp}[+X(E)t], \eqno(11)$$
where $X$ is comparable to, but slightly larger than, the Liapounov exponent
${\chi}$.

This exponential divergence is easy to understand\cite{Pf} and the
functional dependence on ${\Theta}$, ${\eta}$, and ${\chi}$ can actually
be derived theoretically.\cite{KW} The obvious point is that
the unperturbed deterministic trajectory is an unstable stochastic orbit, so
that even the tiniest perturbing influences will tend to grow exponentially.
The average rate of instability is given by ${\chi}$ and, as such, moments
like ${\delta}r_{rms}$ should grow at a rate $X{\;}{\sim}{\;}{\chi}$. That $X$
is slightly larger than ${\chi}$ is a reflection of the fact that, for
different noisy realisations, one sees somewhat different local Liapounov
exponents, and that the total ${\delta}r_{rms}$ will be dominated by those
noisy realisations for which the rate of instability is above average.

This argument might suggest that, although the unperturbed and perturbed
trajectories exhibit a rapid pointwise divergence, their statistical
properties should be virtually identical. Specifically, one might anticipate
that, on short times, the only effect of the friction and noise is to
continually displace the trajectory from one stochastic orbit to another with
essentially the same statistical properties. This, however, is false. Under
certain circumstances, even very weak friction and noise can also alter the
statistical properties of ensembles of stochastic orbits on relatively
short times ${\ll}{\;}100t_{cr}$. Specifically, one observes that such
perturbing influences can dramatically accelerate the rate of penetration
through cantori by providing an additional source of {\it extrinsic diffusion}.

Provided that the friction and noise are sufficiently weak, on short timescales
the energy $E$ is almost conserved, so that one can speak meaningfully of an
evolution restricted to an ``almost constant energy hypersurface.'' Suppose
now that, for some energy $E$, this hypersurface contains large measures of
both sticky and filling stochastic orbits. If, for this energy, the
near-invariant distribution described in Section $4$ is evolved into the
future, allowing for even very weak friction and noise, one then observes a
rapid ($t{\;}{\ll}{\;}100t_{cr}$)  systematic evolution towards a new noisy
near-invariant distribution which is (1) quite different from the
deterministic near-invariant distribution and (2) much closer to the true
deterministic invariant distribution. In this sense, it appears that, on
timescales short compared the timescale $t_{R}$ on which the system would
evolve towards a thermal state, the principal effect of the friction and noise
is to {\it accelerate the approach towards a deterministic invariant
distribution which, in the absence of these perturbing influences, would only
have been realised on much longer timescales}.

The key point in all of this is that friction and noise can induce changes in
orbit class, from filling to confined stochastic, and vice versa. Moreover,
when the deterministic invariant distribution contains large measures of both
sticky and filling orbits, such transitions can happen within a time
$t<100t_{cr}$, even for very weak friction and noise. Visual inspection of
${\sim}{\;}2.5\times 10^{4}$ orbits in several different potentials leads to
the following conclusions.

Typically, for a relaxation time $t_{R}$ as long as $10^{12}t_{cr}$, not many
such changes are observed within a time $t{\;}{\sim}{\;}100t_{cr}$. However,
if $t_{R}$ be reduced to a value ${\sim}{\;}10^{9}t_{cr}$, transitions begin
to become more frequent, and, even for $t_{R}$ as large as
${\sim}{\;}10^{6}t_{cr}{\;\,}$, transitions are quite common, occuring for
$>50\%$ of orbits within a time $t{\;}{\sim}{\;}100t_{cr}$. If the amplitude
of the friction and noise are further increased, one finds that, for
$t_{R}{\;}{\sim}{\;}10^{3}t_{cr}{\;\,}$, transitions are so common that the
distinction between filling and confined becomes essentially meaningless. The
distinction between regular and stochastic is more robust. Only for $t_{R}$
as small as ${\sim}{\;}10^{3}t_{cr}$ are significant numbers of transitions
between regular and stochastic orbits observed within a time as short as
$t{\;}{\sim}{\;}100t_{cr}$.

The fact that even very weak friction and noise, with
$t_{R}{\;}{\sim}{\;}10^{6}t_{cr}-10^{9}t_{cr}$, can significantly alter the
statistical properties of ensembles of orbits on timescales
$t<t_{H}{\;}{\sim}{\;}100t_{cr}$ has direct astronomical implications since,
e.g., for galaxies, the timescale $t_{R}$ for discreteness effects, i.e.,
``collisionality'' is ${\sim}{\;}10^{6}t_{cr}$! The natural timescale
associated with external perturbations is less easily estimated, but may well
be even shorter.

To summarise, it is evident that even very weak friction and noise can alter
both the {\it pointwise} and the {\it statistical} properties of stochastic
orbits in a nonintegrable potential on relative short timescales
${\ll}{\;}t_{R}$. In particular, such effects may be manifest already on time
scales much shorter than the time on which numerical errors in a simulation
can accumulate. This fact has direct and immediate implications for the
problem of ``shadowing'' for numerical orbits.\cite{DVA,Bow}
Physicists, mathematicians, and astronomers are often worried\cite{QT,DEW}
about whether numerical simulations performed on
a computer,
which incorporate roundoff and/or truncation error, can correctly shadow the
evolution of some model system described by a simple set of deterministic
differential equations. However, it would also seem relevant\cite{EF}
to worry about whether the ``real world,'' replete with other
sorts of irregularities, can shadow either the model system or its numerical
realisations. In this regard, one final remark is in order: Rather than being
an impediment to realistic modeling, in certain cases numerical noise may
actually be a good thing, in that it may capture, at least qualitatively, some
of the effects of small perturbing influences to which real systems are always
subjected.

\section*{Acknowledgments}

It is a pleasure to acknowledge useful collaborations with Robert A. Abernathy,
Brendan O. Bradley, George Contopoulos, Salman Habib, Eric O`Neill, Haywood
Smith, Jr., Christos V. Siopis, David E. Willmes, and, especially, M. Elaine
Mahon. This research was supported in part by the NSF through PHY92-03333 and
by NASA through the Florida Space Grant Consortium. The simulations reported
herein were facilitated by time made available by the Advanced Computing
Laboratory at Los Alamos (CM5), The Parallel Computing Research Laboratory at
the University of Florida (KSR), and the Northeast Regional Data Center
(Florida) by the IBM Corp.

\end{document}